\documentclass[prd,preprintnumbers,twocolumn,eqsecnum,floatfix,a4paper,nofootinbib,superscriptaddress]{revtex4-1}
\usepackage{color}
\usepackage{calc}
\usepackage{amsmath,amssymb,graphicx}
\usepackage{amssymb,amsmath}
\usepackage{mathtools}
\usepackage{bm}
\usepackage{microtype}
\usepackage{booktabs}
\usepackage{times}
\usepackage[varg]{txfonts}
\usepackage[colorlinks, pdfborder={0 0 0}]{hyperref}
\usepackage[utf8]{inputenc}
\definecolor{LinkColor}{rgb}{0.75, 0, 0}
\definecolor{CiteColor}{rgb}{0, 0.5, 0.5}
\definecolor{UrlColor}{rgb}{0, 0, 0.75}
\hypersetup{linkcolor=LinkColor}
\hypersetup{citecolor=CiteColor}
\hypersetup{urlcolor=UrlColor}
\allowdisplaybreaks
\DeclareFontFamily{OT1}{pzc}{}
\DeclareFontShape{OT1}{pzc}{m}{it}{<-> s * [1.10] pzcmi7t}{}
\DeclareMathAlphabet{\mathpzc}{OT1}{pzc}{m}{it}

\newcommand{\h}{\mathbf{h}}

\newcommand{\B}{\mathpzc{B}}

\newcommand{\fqnm}{f}
\newcommand{\sigmaqnm}{\sigma}

\makeatletter

\newcommand{\rom}[1]{\expandafter\@slowromancap\romannumeral #1@}
\makeatother

\begin{document}

\newcommand{\be}{\begin{equation}}
\newcommand{\ee}{\end{equation}}
\def\bea{\begin{eqnarray}}
\def\eea{\end{eqnarray}}
\newcommand{\etal}{\emph{et al}}

\title{Including mode mixing in a higher-multipole model for gravitational waveforms\\from nonspinning black-hole binaries}
\author{Ajit Kumar Mehta}
\affiliation{International Centre for Theoretical Sciences, Tata Institute of Fundamental Research, Bangalore 560089, India}
\author{Praveer Tiwari}
\affiliation{International Centre for Theoretical Sciences, Tata Institute of Fundamental Research, Bangalore 560089, India}
\affiliation{Department of Physics, Indian Institute of Science, Bangalore 560012, India}
\affiliation{Department of Physics \& Astronomy, Washington State University, Pullman, WA 99164, USA}
\author{Nathan~K.~Johnson-McDaniel}
\affiliation{International Centre for Theoretical Sciences, Tata Institute of Fundamental Research, Bangalore 560089, India}
\affiliation{Department of Applied Mathematics and Theoretical Physics, Centre for Mathematical Sciences, University of Cambridge, Wilberforce Road, Cambridge, CB3 0WA, UK}
\author{Chandra Kant Mishra}
\affiliation{Indian Institute of Technology Madras, Chennai 600036, India}
\author{Vijay Varma}
\affiliation{Theoretical Astrophysics, 350-17, California Institute of Technology, Pasadena, CA 91125, USA}
\affiliation{International Centre for Theoretical Sciences, Tata Institute of Fundamental Research, Bangalore 560089, India}
\author{Parameswaran~Ajith}
\affiliation{International Centre for Theoretical Sciences, Tata Institute of Fundamental Research, Bangalore 560089, India}
\affiliation{Canadian Institute for Advanced Research, CIFAR Azrieli Global Scholar, MaRS Centre, West Tower, 661 University Ave., Suite 505, Toronto, ON M5G 1M1, Canada}

\begin{abstract}
As gravitational-wave (GW) observations of binary black holes are becoming a precision tool for physics and astronomy, several subdominant effects in the GW signals need to be accurately modeled. Previous studies have shown that neglecting subdominant modes in the GW templates causes an unacceptable loss in detection efficiency and large systematic errors in the estimated parameters for binaries with large mass ratios. Our recent work [Mehta~\emph{et al.}, Phys.~Rev.~D {\bf{96}}, 124010 (2017)] constructed a phenomenological gravitational waveform family for nonspinning black-hole binaries that includes subdominant spherical harmonic modes $(\ell = 2, m = \pm 1)$, $(\ell = 3, m = \pm 3)$, and  $(\ell = 4, m = \pm 4)$ in addition to the dominant quadrupole mode, $(\ell = 2, m=\pm 2)$. In this article, we construct analytical models for the ($\ell = 3, m = \pm 2$) and ($\ell = 4, m = \pm 3$) modes and include them in the existing waveform family. Accurate modeling of these modes is complicated by the mixing of multiple spheroidal harmonic modes. We develop a method for accurately modeling the effect of mode mixing, thus producing an analytical waveform family that has faithfulness $> 99.6\%$.
\end{abstract}
\maketitle
\section{Introduction}
The detection of compact binary coalescences is now commonplace for the Advanced LIGO~\cite{TheLIGOScientific:2014jea} and Advanced Virgo~\cite{TheVirgo:2014hva} detectors, and they have now produced their first catalogue of such detections~\cite{LIGOScientific:2018mvr}. The accurate extraction of the parameters and hence the science output from these events depends on the accurate modeling of the gravitational waves (GWs) from such sources. Detection of GWs from compact binaries primarily relies on the method of matched filtering, which requires hundreds of thousands of signal templates to be compared against the data (e.g.,~\cite{DalCanton:2017ala,Mukherjee:2018yra}). Inference of source parameters from observed signals also relies on comparing the data with theoretical waveform templates~\cite{LSC_2016paramest}. Although numerical relativity (NR) provides the most accurate template waveforms, the large computational cost and sparse parameter space coverage of the NR simulations make the direct implementation of NR waveforms in GW data analysis challenging (see~\cite{Abbott:2016apu, Lange:2017wki} for some recent work in this direction). Over the past decade, there has thus been a considerable effort devoted to developing quick-to-evaluate accurate waveform models for the detection and parameter estimation of GWs from the inspiral, merger, and ringdown of binary black holes, e.g.,~\cite{Buonanno:1998gg,Buonanno:2000ef,Pan:2010hz,Pan:2011gk,Pan:2013rra,Damour:2009kr,Damour:2008gu,Damour:2012ky,Ajith:2007kx,Ajith:2009bn,Santamaria:2010yb,Hannam:2013oca,Husa_2016IMRPhenomD,Khan_2016IMRPhenomD,Taracchini:2013,Babak:2016tgq,Bohe:2016gbl,Nagar:2018zoe,Khan:2018fmp,McWilliams:2018ztb}. 

Most of these (semi) analytical waveforms contain only the dominant multipoles (quadrupole) of the gravitational radiation, though higher multipole models are now starting to be developed~\cite{Pan:2011gk,Blackman:2015pia,Blackman:2017pcm, London:2017bcn, Cotesta:2018fcv, Varma:2018mmi, Nagar:2019wds, Varma:2019csw}. Studies show that neglecting the higher modes can result in a considerable reduction in the sensitivity of searches for high-mass, higher-mass-ratio binary black holes~\cite{Varma:2016dnf, CalderonBustillo:2016hm, Varma:2014hm, Harry:2017weg}. Neglecting these modes can also lead to systematic biases in the parameter estimation of LIGO events from binaries with large mass ratios or high inclination angles, thus biasing our inference of the astrophysical properties of the sources~\cite{Varma:2016dnf, Varma:2014hm, Abbott:2016wiq}. The inclusion of higher multipoles is also expected to provide several other advantages, such as improvements in the precision of parameters extracted from the data~\cite{Sintes:1999cg,VanDenBroeck:2006ar,Arun:2007hu,Trias:2008pu,Arun:2008zn,Graff:2015bba,Lange:2017wki,OShaughnessy:2017tak,London:2017bcn,Kumar:2018hml} and in the accuracy of various observational tests of GR~\cite{Pang:2018hjb}, the detection of GW memory~\cite{Lasky:2016knh,Talbot:2018sgr}, etc.

In this paper, we extend our previous higher multipole waveform family for nonspinning binary black holes~\cite{Mehta:2017jpq} to include some additional subdominant spherical harmonic modes ($\ell=3, m=2$ and $\ell=4, m=3$) which have a more complicated behavior in their post-merger part of the waveforms due to an effect known as \emph{mode mixing}.\footnote{Since we are considering nonprecessing binaries, we will only refer to the modes with positive $m$ explicitly, as the modes with negative $m$ can be obtained from the positive $m$ modes by symmetry. Additionally, we will refer to the modes using just $\ell m$ for the remainder of the paper.} As a consequence, they posses some unusual bumps in the post-merger part which make it difficult to model them accurately. The dominant cause for mode mixing is the mismatch between angular basis functions used in NR (i.e., spin $-2$ weighted \emph{spherical} harmonics) and in Kerr black hole perturbation theory (i.e., spin $-2$ weighted \emph{spheroidal} harmonics)~\cite{Kelly:2012nd}.\footnote{See~\cite{Buonanno:2006ui} for an initial study of this effect and~\cite{London:2014cma} for a further study of mode mixing in numerical relativity waveforms.}
As GW observations are entering a regime of precision astronomy, such as precision tests of general relativity (GR)~\cite{O2:testingGR,Ghosh:2017gfp,Meidam:2017dgf,Dhanpal:2018ufk}, modeling of such subtle effects in the waveforms becomes important. 

We introduce a method for approximately extracting the unmixed spheroidal harmonic modes from the spherical harmonic modes (the quasi-normal modes separate in the spheroidal basis). We then model these unmixed modes using suitable phenomenological functions motivated by black hole perturbation theory and finally reintroduce the mixing to obtain the model for the spherical harmonic modes. The resulting waveform model is highly faithful (faithfulness $> 99.6\%$) and fast to evaluate. 

We describe our method for removing the mode mixing approximately and modeling the resulting ``unmixed'' modes in Sec.~\ref{sec:waveform_model}, where we also describe how we add the mode mixing back in to the final model and test its accuracy by computing matches with hybrid waveforms. We summarize and conclude in Sec.~\ref{sec:summ_concl}. Additionally, we list the waveforms used for calibration and validation in Appendix~\ref{sec:SXS} and give some additional plots in Appendix~\ref{sec:lower_mass_rat_plot}. Throughout the paper we denote the binary's total mass by $M$. We also denote the real and imaginary parts of quantities by a superscript R and I, respectively.

\begin{figure*}[tbh] 
	\begin{center}
		\includegraphics[height=2.9in]{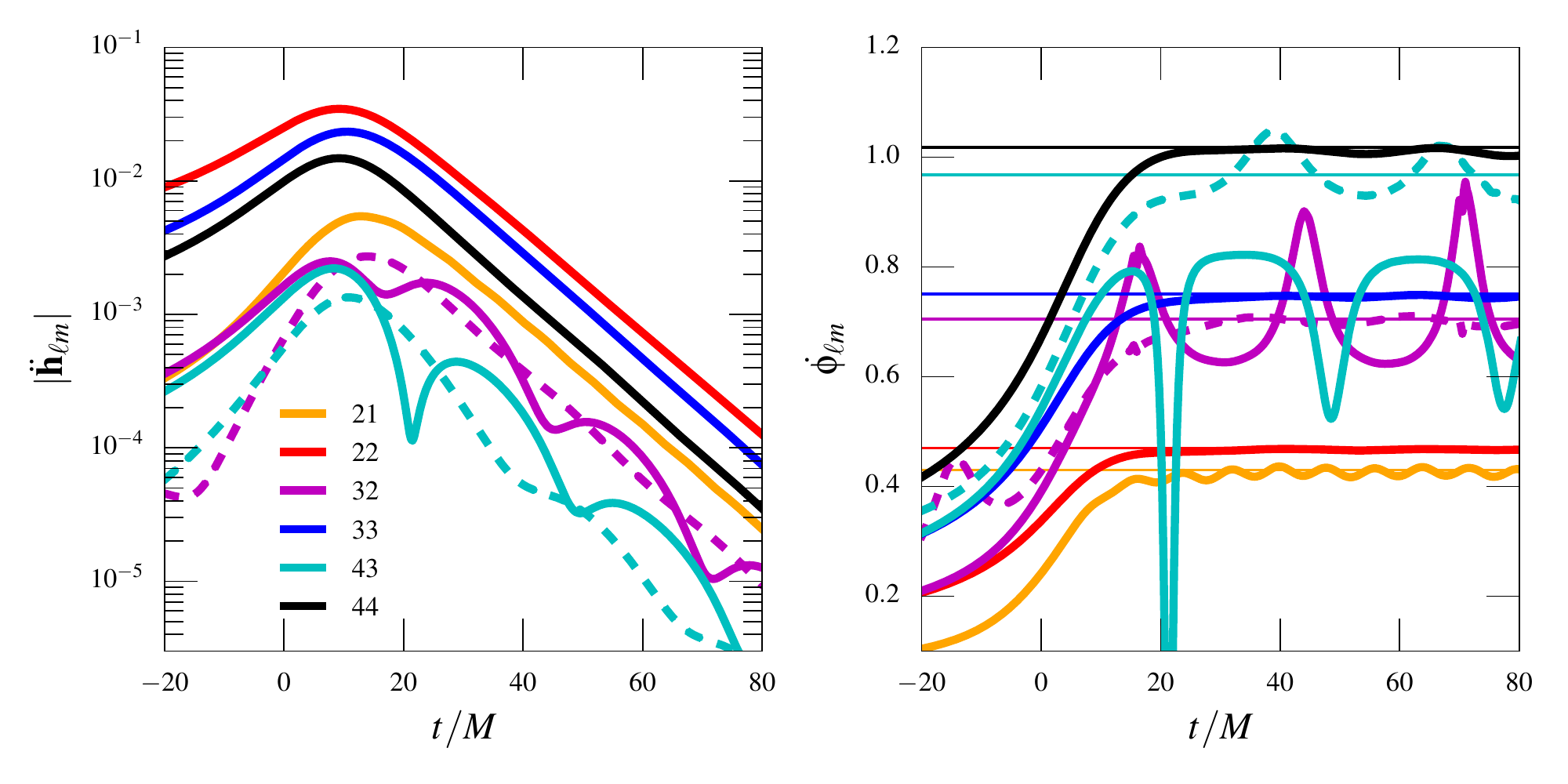}
	\end{center}
	\vspace*{-3mm}
	\caption{\emph{Left panel:} Amplitude of the second time derivative of different spherical harmonic modes ${}^\text{Y}\ddot{\h}_{\ell m}(t)$ (solid lines) from a nonspinning binary with  mass ratio $q=4$. Time $t=0$ corresponds to the peak amplitude of $22$ mode. Note the oscillations in the $32$ and $43$ modes for $t > 0$, due to the mixing of multiple spheroidal harmonic modes. The dashed lines show the amplitude of the second time derivative of the \emph{spheroidal} harmonic modes ${}^\text{S}\ddot{\h}_{\ell m 0}(t)$ for $\ell m \in \{32, 43\}$ constructed using the prescription presented in Sec.~\ref{sec:mode_mixing_removal}, which are better behaved in the ringdown regime ($t > 0$). \emph{Right panel:} The instantaneous frequency $\dot {\phi}_{\ell m}(t)$ of the second time derivatives of the spherical (solid lines) and spheroidal (dashed lines) modes. The horizontal lines show the quasi-normal-mode frequencies of different modes. Note that the $32$ and $43$ spherical harmonic modes' frequencies (solid lines) do not approach the corresponding quasi-normal-mode frequencies, while the spheroidal harmonic modes' frequencies (dashed lines) do. Note also that the oscillations in the instantaneous frequency of the $21$ mode are not due to mixing from the $31$ mode, which has an indistinguishable effect in this plot. These are rather likely due to the interference between positive and negative frequency QNMs, which is found to be particularly prominent for the $m = 1$ modes in~\cite{Bernuzzi:2010ty}; see also~\cite{Nagar:2019wds}.} 
	\label{fig:amp_comp}
\end{figure*}

\begin{figure}[tbh] 
	\begin{center}
		\includegraphics[height=2.9in]{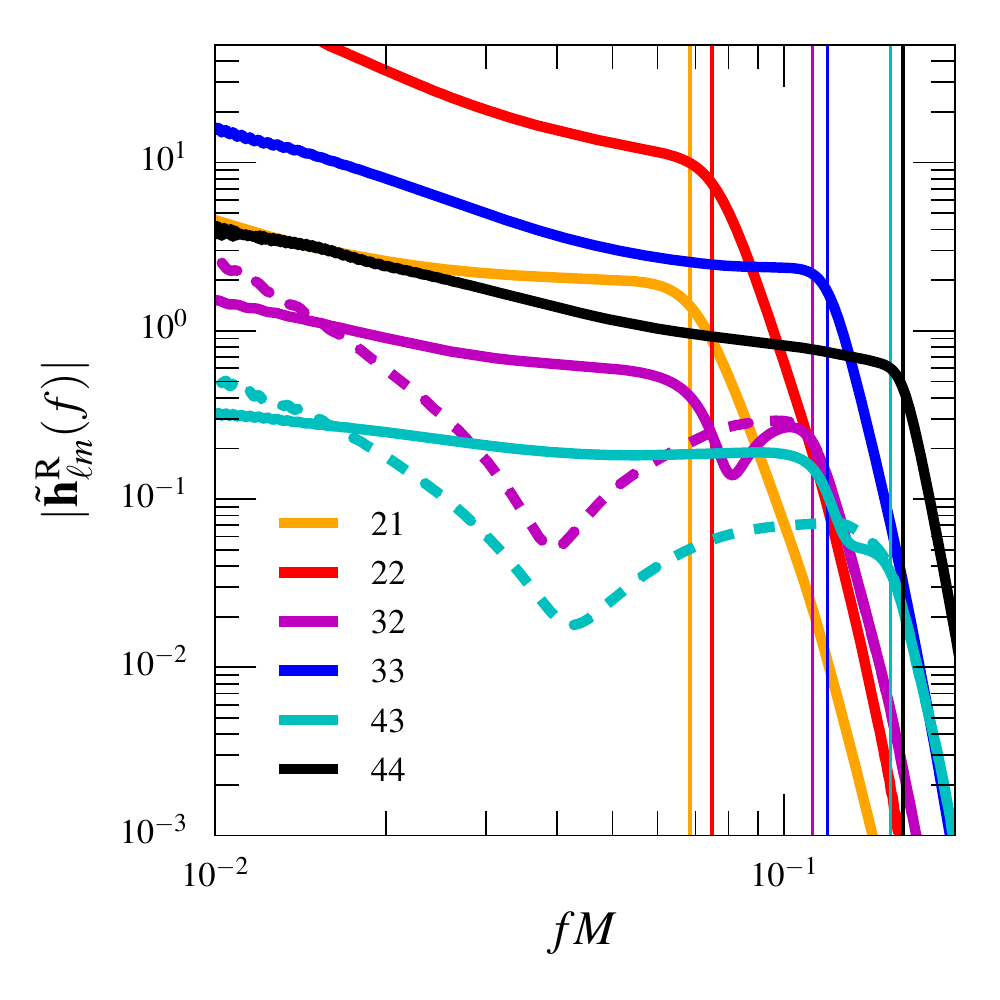}
	\end{center}
	\vspace*{-3mm}
	\caption{Fourier domain amplitude $|\tilde{\h}^\text{R}_{\ell m}(f)|$ of the spherical (solid) and spheroidal (dashed) harmonic modes from a nonspinning binary with  mass ratio $q=4$. The vertical line with the corresponding color represents  $f^\text{QNM}_{\ell m}$.}
	\label{fig:amp_comp_fd}
\end{figure}

\section{An improved waveform model for subdominant modes}
\label{sec:waveform_model}

\subsection{Mixing of spherical and spheroidal harmonic modes}
\label{sec:mode_mixing}

The two polarizations $h_+(t)$ and $h_\times(t)$ of GWs can be expressed as a complex waveform $\h(t) := h_+(t) - \mathrm{i} \, h_\times(t)$. It is convenient to expand this in terms of the spin $-2$ weighted spherical harmonics so that the radiation along any direction $(\iota, \varphi_0)$ in the source frame can be expressed as 
\begin{equation}
\h(t; \iota, \varphi_0) = \sum_{\ell \geq 2}\sum_{|m| \leq \ell} Y_{\ell m}(\iota, \varphi_0) \, {}^\text{Y}\h_{\ell m}(t).
\label{eq:hoft_sphericalH}
\end{equation}
The spherical harmonic modes ${}^\text{Y}\h_{\ell m}(t) = A_{\ell m}(t) \, e^{\mathrm{i} \, \phi_{\ell m}(t)}$ are purely functions of the intrinsic parameters of the system (such as the masses and spins of the binary), while all the angular dependence is captured by the spherical harmonic basis functions $Y_{\ell m}(\iota, \varphi_0)$. Here, by convention, the polar angle $\iota$ is measured with respect to the orbital angular momentum of the binary. The leading contribution to $\h(t; \iota, \varphi_0)$ comes from the quadrupolar $22$ modes. 

We construct spherical harmonic modes of hybrid waveforms for different modes using the method described in~\cite{Varma:2014hm,Mehta:2017jpq}. Figure~\ref{fig:amp_comp} shows the amplitude (solid lines in left panel) and instantaneous frequency (solid lines in right panel) of the second time derivative of different spherical harmonic modes of the hybrid waveforms with mass ratio $q = 4$.\footnote{We consider the second time derivative of $\h$ (i.e., the Weyl scalar $\psi_4$) here instead of $\h$ itself in order to give a cleaner illustration. If we make the same plot using $\h$, we find additional oscillations, even in modes that are not expected to have significant mode mixing. These oscillations appear to be due primarily to additional constant and linear terms in $\h$ that are removed by taking the time derivatives. Taking a single time derivative of $\h$ (i.e., considering the Bondi news) removes most of the oscillations, but taking a second time derivative removes some remaining oscillations. Since we are concerned with removing the mode mixing in the frequency domain, where the time derivatives correspond to a multiplicative factor, there is nothing lost in illustrating the mode mixing removal in the time domain using $\psi_4$. We compute the second time derivative by second-order accurate finite differencing.} We note that the $22$, $33$, $44$, and $21$ modes, for which an analytical phenomenological model was presented in \cite{Mehta:2017jpq}, have smoothly varying amplitude and frequency. On the other hand, the $32$ and $43$ modes have some bumps in the post-merger regime $(t > 0)$. The unusual behavior of these modes is attributed to what is known as \emph{mode-mixing}, where multiple \emph{spheroidal} harmonic modes are getting mixed in one \emph{spherical} harmonic mode. The prime cause of the mode mixing is the mismatch between the angular basis that is used in NR simulations to extract waveforms (spherical harmonics) and the one that is used to separate the Teukolsky equations in Kerr black hole perturbation theory (spheroidal harmonics)~\cite{Kelly:2012nd}.

The mixing of multiple spheroidal harmonic modes creates multiple frequencies in the ringdown waveform that makes it hard to model them using simple analytical functions. Figure~\ref{fig:amp_comp_fd} shows an example of the Fourier domain amplitude $|{}^\text{Y}\tilde{\h}^\text{R}_{\ell m}(f)|$ of different hybrid modes --- note the non-monotonic behavior seen in the higher frequencies of the $32$ and $43$ modes. These modes were thus left out in the phenomenological model presented in~\cite{Mehta:2017jpq}. In this paper, we present a phenomenological model for the $32$ and $43$ modes. Our approach is to subtract the effect of mode mixing from these modes which allows us to model these ``unmixed'' modes using methods that were used earlier, and then reintroduce the effects of mode mixing to obtain the final model.

\begin{figure}[htb] 
	\begin{center}
	\includegraphics[height=2.95in]{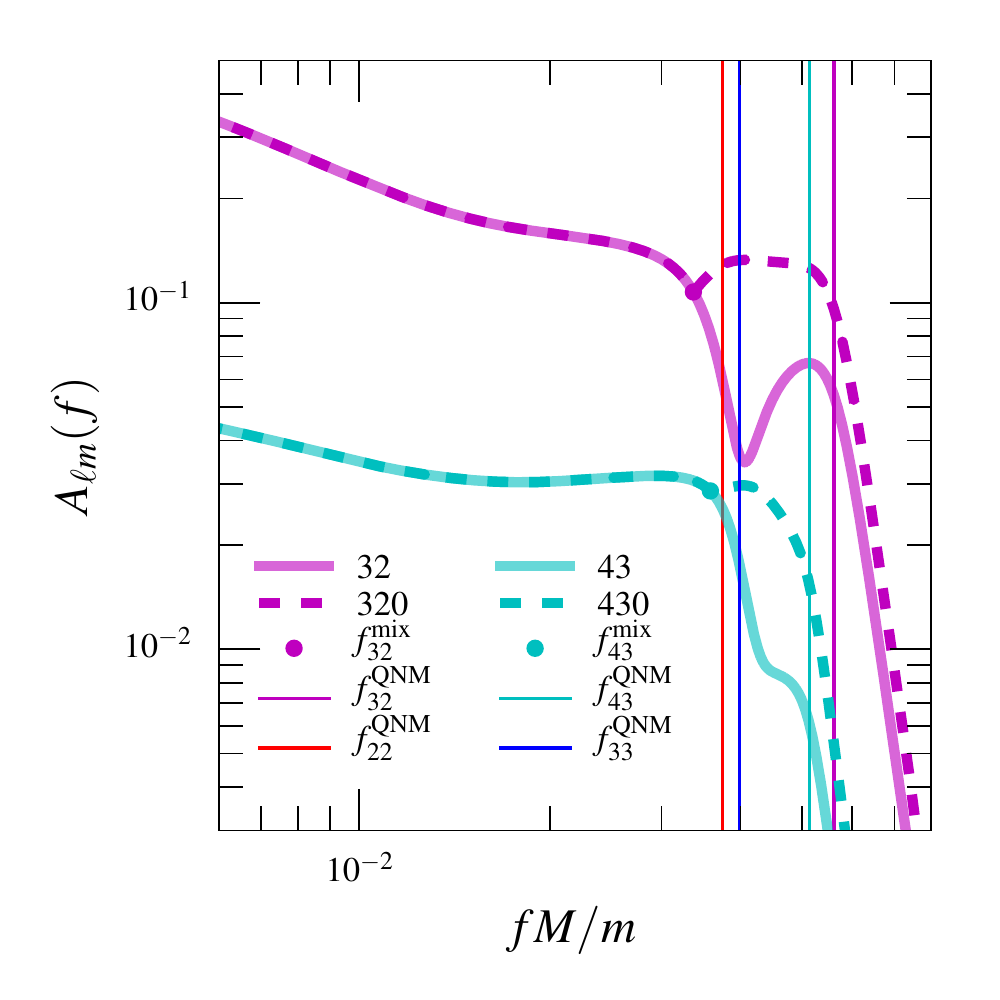}
	\end{center}
	\vspace*{-3mm}
	\caption{The amplitude of  mixed and unmixed modes  as a function of frequency for  mass ratio $q=4$. The dashed lines represent the amplitude of unmixed modes. The $43$ mode has been scaled appropriately to avoid overlap with the $32$ mode.}
	\label{fig:mode_mixing_freq}
\end{figure}

\begin{figure*}[htb] \begin{center}
		\includegraphics[height=2.95in]{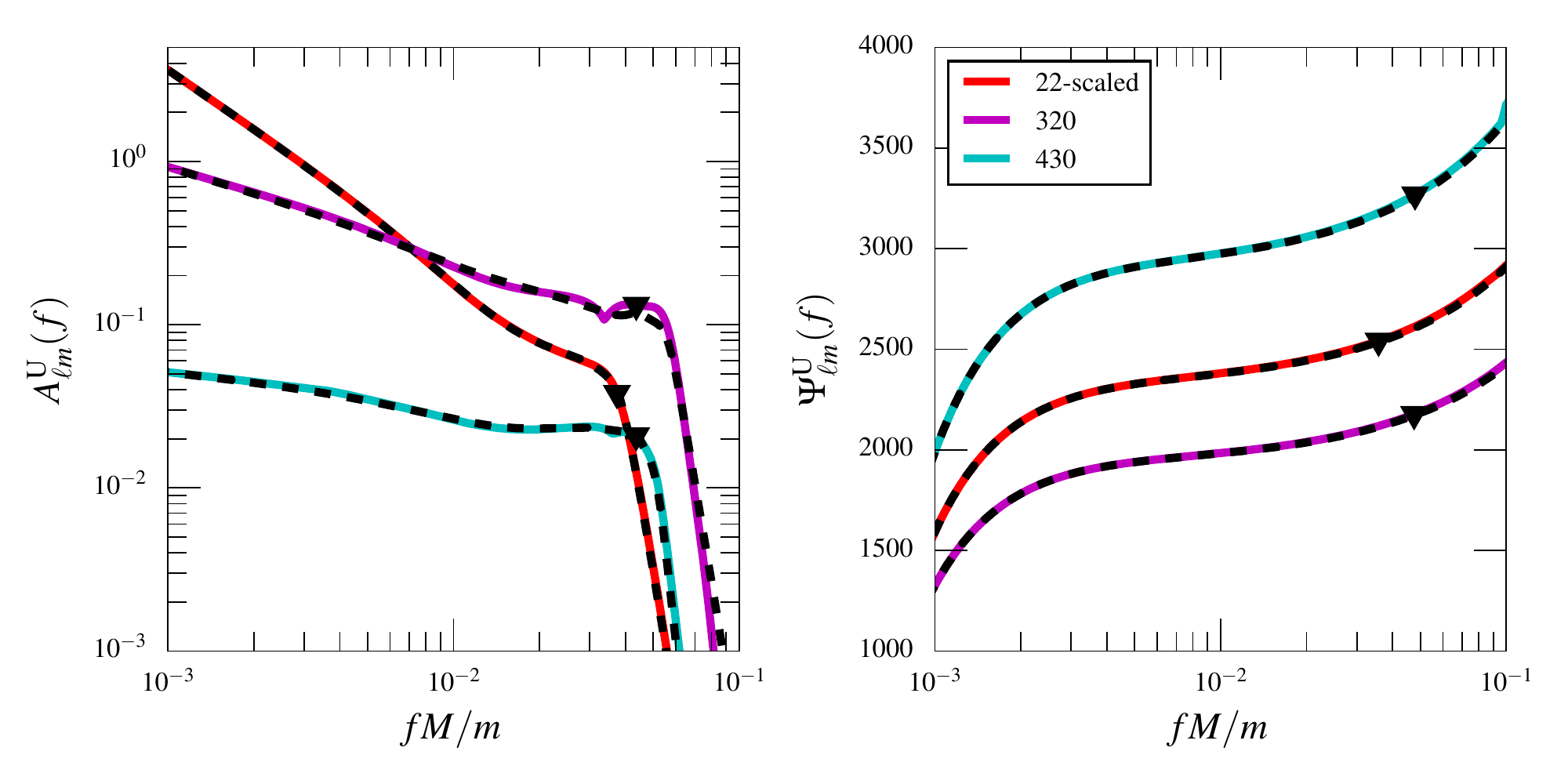}
		\vspace*{-2mm}
		\caption{Comparison between the amplitude (left panel) and phase (right panel) of the unmixed modes for the hybrid and analytical model waveforms for  mass ratio $q = 4$. In each plot, the solid lines correspond to the unmixed modes and the dashed lines correspond to the analytical model waveforms for the same mode. The black triangles represent the transition frequency from inspiral-merger to ringdown as defined in Eqs.~\eqref{eq:phenom_ampl_model} and~\eqref{eq:phenom_phase_model}, i.e., $f^{\rm A}_{\ell m}$ and $f^{\rm P}_{\ell m}$. 
			The amplitude and phase of the $22$ mode have been scaled appropriately to make them fit inside the figure.}
		\label{fig:comp_waveform}
\end{center}\end{figure*}

\subsection{Removal of mode mixing from the 32 and 43 modes}
\label{sec:mode_mixing_removal}

The binary merger produces a perturbed black hole which settles into a stationary Kerr black hole. Teukolsky's solution for GWs from a perturbed Kerr black hole has a natural decomposition in spin $-2$ weighted \emph{spheroidal} harmonics,  $S_{\ell mn} \equiv S_{\ell m}(a_f \omega_{\ell mn})$ associated with quasinormal mode (QNM) frequencies $\omega_{\ell mn}$, where $M_fa_f$ is the spin angular momentum of the final black hole (of mass $M_f$). See, e.g.,~\cite{Berti:2005gp} for information about the properties of these functions. Thus, GW polarizations from the ringdown can be written as
\begin{equation}
\h(t; \iota, \varphi_0) =  \sum_{\ell' \geq 2} \sum_{|m| \leq \ell'} \sum_{n \geq 0} S_{\ell' m n}(\iota, \varphi_0) ~ {}^\text{S}\h_{\ell^\prime m n}(t). 
\label{eq:spheroidal_decomp}
\end{equation}
Here the overtone index $n$ measures the magnitude of the imaginary part of the quasinormal mode frequencies $\omega_{\ell mn}$. Note that the spheroidal harmonic basis functions  $S_{\ell^{\prime}mn}$ can be expressed in terms of (spin $-2$ weighted) spherical harmonics $Y_{\ell m}$ as
\begin{equation}
S_{\ell^{\prime}mn} = \sum_{\ell\geq|m|} \mu_{m \ell \ell^{\prime}n}^{*} Y_{\ell m},
\end{equation}
where $\mu_{m \ell \ell^{\prime}n}$ are mixing coefficients which can be computed simply using the fits provided by Berti and Klein~\cite{Berti:2014fga} (there are more complicated fits given in~\cite{London:2018nxs}) and the star denotes the complex conjugate.\footnote{We actually substitute $\mu_{m \ell \ell^{\prime}n} \to (-1)^{\ell + \ell'}\mu_{m \ell \ell^{\prime}n}$, where the prefactor corrects for the difference in the sign convention for spin-weighted spherical harmonics that we use---the same convention as~\cite{Ajith:2007jx}, which is also the one used in the SpEC code~\cite{Boyle_PC}---and the one used by Berti and Klein. There is an additional factor of $(-1)^m$ that we neglect, as it is fixed for each mode we consider (including its mixed modes).} By inserting this expansion in Eq.~(\ref{eq:spheroidal_decomp}), we have
\begin{equation}
\h(t; \iota, \varphi_0) =  \sum_{\ell' \geq 2} \sum_{|m| \leq \ell'} \sum_{n \geq 0} \sum_{\ell\geq|m|} \mu_{m \ell \ell^{\prime}n}^{*} Y_{\ell m}(\iota, \varphi) ~    {}^\text{S}\h_{\ell^\prime m n}(t). 
\end{equation}
Comparing this with Eq.~\eqref{eq:hoft_sphericalH}, we get 
\begin{equation}
{}^\text{Y}\h_{\ell m}(t) = \sum_{\ell^{\prime}\geq|m|}\sum_{n\geq0} {}^\text{S}\h_{\ell^{\prime}mn}(t) \, \mu_{m \ell \ell^{\prime}n}^{*}. 
\label{eq:spherical_to_spheroidal}
\end{equation}
Thus, spherical harmonic modes of the hybrid waveforms can be written in terms of the spheroidal harmonic modes. From inspection of the different spherical harmonic modes of the NR data, we get an understanding of the relative amplitudes of these modes (see, e.g., Fig.~1 in both~\cite{Pan:2011gk} and~\cite{Cotesta:2018fcv}). We thus make the following approximations when removing the mode mixing: 
\begin{itemize}
\item The amplitudes of the higher spheroidal overtones are negligible because their damping times are factors of $\gtrsim 3$ smaller than those of the leading overtone $n = 0$. Hence we will only consider mixing from the leading overtone. 
\item For $\ell = m$ spherical modes, the mixing contribution from any mode except the $\ell \ell 0$
spheroidal mode is negligible. 
\item For a general $\ell m$ spherical mode, contribution from spheroidal modes with $\ell^{\prime} >\ell$ is negligible, since the higher mode amplitudes are much smaller than the $\ell m 0$ spheroidal mode, and they are also multiplied by the mixing coefficient which is already small. 
\end{itemize}
As a result of these approximations, a particular $\ell m$ spherical mode will have contribution from spheroidal modes $\ell^{\prime}m0$ with $\ell^{\prime}\leq \ell$ (and the obvious restriction of $\ell^{\prime} \geq |m|$). We thus have
\begin{equation}
{}^\text{Y}\h_{\ell m}(t) \simeq \sum_{\ell^{\prime}\leq \ell} {}^\text{S}\h_{\ell^{\prime}m0}(t) \, \mu_{m \ell \ell^{\prime}0}^{*}. 
\end{equation}
To determine the spheroidal modes ${}^\text{S}\h_{\ell^{\prime}m0}(t)$ from the spherical modes ${}^\text{Y}\h_{\ell m}(t)$, we observe that it is a perfectly determined system of coupled equations when we consider different $\ell m$ spherical modes. To be specific, we compute the following spheroidal modes:\footnote{The same procedure also works for modes with $m \leq \ell - 2$ that have three or more spheroidal modes mixed into the spherical mode in our approximation, e.g., the 42 mode studied in~\cite{Praveer_thesis}. However, this mode has a small enough amplitude that we do not include it in the present study.}
\begin{subequations}
\label{eq:mode_remov}
\begin{align}
{}^\text{S}\h_{320}(t) &\simeq \dfrac{{}^\text{Y}\h_{32}(t) - {}^\text{Y}\h_{22}(t) \mu_{2320}^{*}/\mu_{2220}^{*}}{\mu_{2330}^{*}}, \label{eq:mode_remov_a}\\ 
{}^\text{S}\h_{430}(t) &\simeq \dfrac{{}^\text{Y}\h_{43}(t) - {}^\text{Y}\h_{33}(t) \mu_{3430}^{*}/\mu_{3330}^{*}}{\mu_{3440}^{*}}. 
\end{align}
\end{subequations}
These spheroidal harmonic modes for a binary with $q = 4$ are shown as dashed lines in Fig.~\ref{fig:amp_comp} (as discussed there, we plot the second time derivatives to give a cleaner illustration). It can be seen that the amplitude oscillations seen in the spherical modes (solid lines) are largely absent in the spheroidal modes (dashed lines). In addition, the instantaneous frequency (right panel) of the spheroidal modes approaches the corresponding quasi-normal-mode frequency. 



We can also convert Eqs.~\eqref{eq:mode_remov} into the frequency domain, so that we can remove the mode mixing from the frequency domain waveforms. Here we want to compute the Fourier transforms of the real and imaginary parts separately, since in this nonprecessing case we can focus on just modeling the real part, and the imaginary part can be obtained from the real part by a phase shift of $\pi/2$. However, we give the expression for the imaginary part as well, for completeness. A straightforward calculation, i.e., taking the real and imaginary parts of Eq.~\eqref{eq:mode_remov_a} and expressing them in the frequency domain, gives us the following form for the $32$ mode:
\begin{equation}
\begin{split}
{}^\text{S}\tilde{\h}_{320}^{\mathrm{R}}(f)  &\simeq (\alpha_{1} \mu_{2330}^{\mathrm{R}} - \alpha_{2} \mu_{2330}^{\mathrm{I}})/|\mu_{2330}|^{2}, \\[0.5ex]
{}^\text{S}\tilde{\h}_{320}^{\mathrm{I}}(f)  &\simeq (\alpha_{2} \mu_{2330}^{\mathrm{R}} + \alpha_{1} \mu_{2330}^{\mathrm{I}} )/|\mu_{2330}|^{2},
\label{eq:freq_mod_remov}
\end{split}
\end{equation}
where 
\begin{equation}
\begin{split}
\alpha_{1} &:= {}^\text{Y}\tilde{\h}_{32}^{\mathrm{R}}(f) - \bigg(  {}^\text{Y}\tilde{\h}_{22}^{\mathrm{R}}(f) \rho_{2320}^{\mathrm{R}} + {}^\text{Y} \tilde{\h}_{22}^{\mathrm{I}}(f) \rho_{2320}^{\mathrm{I}}  \bigg), \\
\alpha_{2} &:= {}^\text{Y}\tilde{\h}_{32}^{\mathrm{I}}(f) + \bigg( {}^\text{Y}\tilde{\h}_{22}^{\mathrm{R}}(f) \rho_{2320}^{\mathrm{I}} - {}^\text{Y}\tilde{\h}_{22}^{\mathrm{I}}(f) \rho_{2320}^{\mathrm{R}}   \bigg).
\end{split}
\end{equation}
Here $\rho_{2320} := \mu_{2320}/\mu_{2220}$ and ${}^\text{Y}\tilde{\h}^\mathrm{R}_{\ell m}(f)$, ${}^\text{Y}\tilde{\h}^\mathrm{I}_{\ell m}(f)$ are the Fourier transforms of the real and imaginary parts of ${}^\text{Y}\h_{\ell m}(t)$, respectively. The expressions for the $43$ mode are analogous.

The amplitude $|{}^\text{Y}\tilde{\h}_{\ell m}^{\mathrm{R}}(f)|$ in the Fourier domain is shown for the $32$ and $43$ modes in Fig.~\ref{fig:mode_mixing_freq} (lighter shades). There are clearly two features in the $32$ mode at close to the QNM frequencies of the $320$ and $220$ modes and similarly for the $43$ mode. Now the ``unmixed'' modes are constructed as follows: 
	\begin{subequations}
	\begin{align}
	A^\text{U}_{\ell m}(f) &:= \begin{dcases}
	|{}^\text{Y}\tilde{\h}_{\ell m}^{\mathrm{R}}(f)|,& \,\, f< f^\text{mix}_{\ell m},\\
		w_{\ell m}^\text{U}|{}^\text{S}\tilde{\h}_{\ell m 0}^{\mathrm{R}}(f)|,& \,\, f \geq f^\text{mix}_{\ell m},
	\end{dcases}\\
	\Psi^\text{U}_{\ell m}(f) &:= \begin{dcases}
	\arg({}^\text{Y}\tilde{\h}_{\ell m}^{\mathrm{R}}(f)),& \,\, f< f^\text{mix}_{\ell m},\\
	\phi_{\ell m}^\text{U}+\arg({}^\text{S}\tilde{\h}_{\ell m 0}^{\mathrm{R}}(f)),& \,\, f \geq f^\text{mix}_{\ell m},
	\end{dcases}
	\end{align}
	\end{subequations}
where $A^\text{U}_{\ell m}(f)$ and $\Psi^\text{U}_{\ell m}(f)$ represent the amplitude and phase of the unmixed modes respectively, while $f^\text{mix}_{\ell m}$ is a transition frequency. The parameters $w_{\ell m}^\text{U}$ and $\phi_{\ell m}^\text{U}$ are determined by demanding the continuity of the amplitude and phase at $f^\text{mix}_{\ell m}$, respectively.

To determine $f^\text{mix}_{\ell m}$, we note that the bump in the amplitude of a certain $\ell m$ spherical harmonic mode due to the mixing of the $(\ell -1) m$ mode always appears at frequencies slightly below the $\ell m$ mode's dominant QNM frequency $f_{\ell m}^\text{QNM}$. When $f^\text{mix}_{\ell m}$ is allowed to be a free parameter, it becomes degenerate with the model parameters [Eq.~\eqref{eq:phenfits_phase_fits}] and thus makes the model fail, i.e., the parameters appearing in Eq.~\eqref{eq:phenfits_phase_fits} do not have a simple dependence on $\eta$. We find that fixing $f^\text{mix}_{\ell m}= 0.9f_{\ell m}^\text{QNM}$ gives good agreement of the model parameters with quadratic functions of $\eta$. In Fig.~\ref{fig:mode_mixing_freq}, we also plot the unmixed modes (dashed lines). The bumps in the amplitudes of the spherical harmonic modes due to mode mixing are significantly suppressed in the unmixed modes.

\begin{figure*}[htb] \begin{center}
		\includegraphics[width=7in]{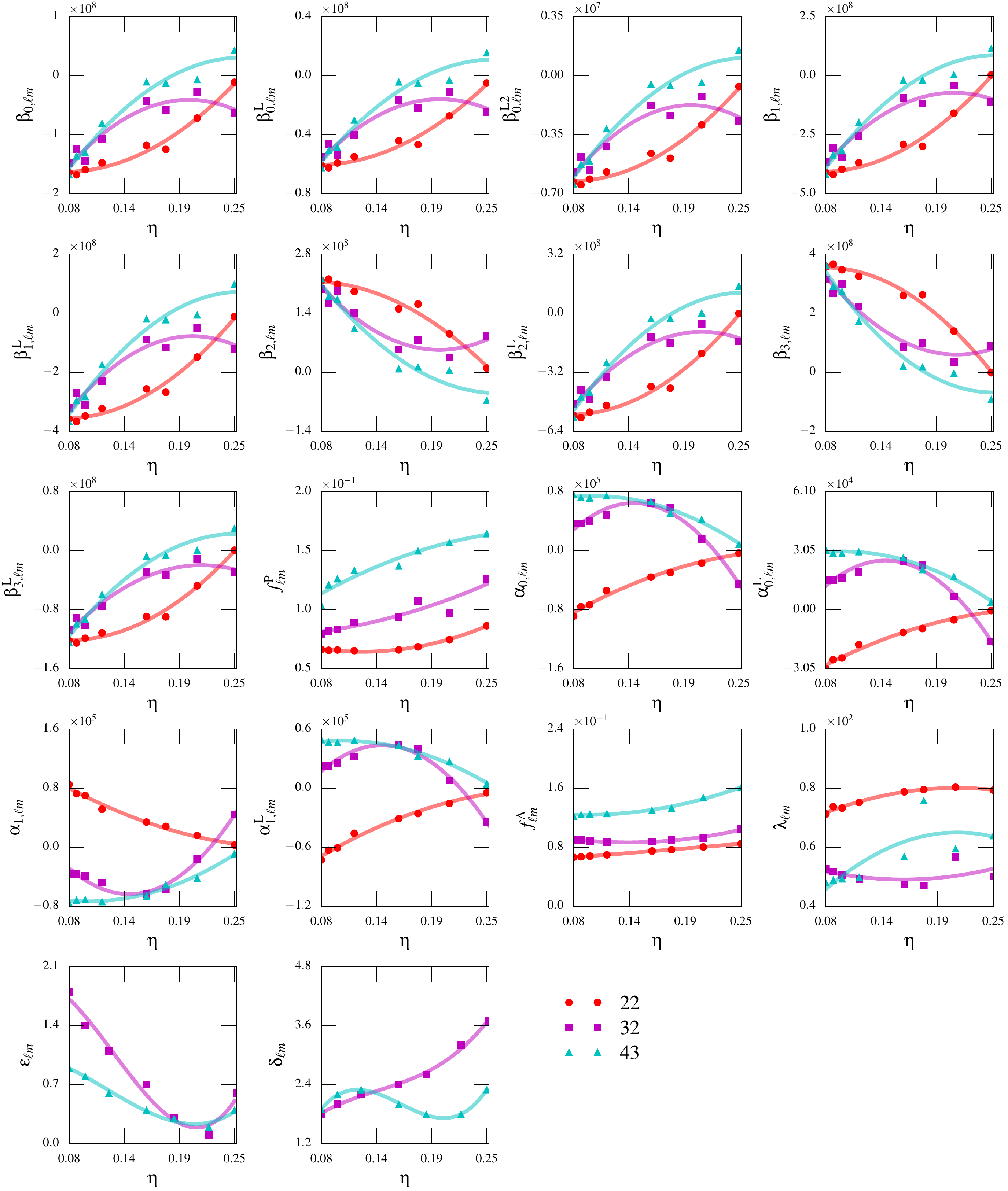}
		\vspace*{-2mm}
		\caption{The estimated values of the phenomenological parameters describing the analytical model waveforms, plotted against the symmetric mass ratio $\eta$. 
		}
		\label{fig:coef_fit}
\end{center}\end{figure*}

\subsection{Construction of the analytical waveform model }
\label{sec:phenom_model}
To construct models for the amplitude $A^\text{U}_{\ell m}(f)$ and phase $\Psi^\text{U}_{\ell m}(f)$, $\ell m \in \{32,43\}$, we follow exactly the same procedure as Sec.~\rom{2}~B of \cite{Mehta:2017jpq}. We calibrate the model to the same hybrid waveforms used to calibrate the model in~\cite{Mehta:2017jpq}. These hybrid waveforms are constructed by matching NR waveforms from the SXS Gravitational Waveform Database~\cite{SXS-Catalog,Mroue:2013xna,Blackman:2015pia,Chu:2015kft}, listed in Appendix~\ref{sec:SXS}, with post-Newtonian (PN)/effective-one-body waveforms, using the procedure described in~\cite{Varma:2014hm,Mehta:2017jpq}. The imaginary part of the unmixed mode (in the time domain) is related to the real part by a phase shift of $\pi/2$, due to the symmetry of nonprecessing binaries. Hence, we only model the Fourier transform of the real part. 
The amplitude model is thus
\begin{equation}
A^\text{U, mod}_{\ell m}(f) = \begin{dcases}
A_{\ell m}^\mathrm{IM}(f),& \,\, f< f^{\rm A}_{\ell m},\\  
A_{\ell m}^\mathrm{RD}(f),& \,\, f \geq f^{\rm A}_{\ell m},
\end{dcases}
\label{eq:phenom_ampl_model}
\end{equation}
where $f^{\rm A}_{\ell m}$  denotes the transition frequency from the inspiral-merger part of the waveform to the ringdown in the amplitude. The  inspiral-merger part is modelled as 
\begin{equation}
A_{\ell m}^\mathrm{IM}(f) = A_{\ell m}^\mathrm{PN}(f) \left[1+ \sum\limits_{k=0}^{k=1}\left(\alpha_{k,\,\ell m}\, +\alpha_{k,\,\ell m}^\text{L}\,\ln v_f \right)\,v_f^{k+8} \right],
\label{eq:phenom_ampl_model_IM}
\end{equation}
where $v_f = (2 \pi M f/m)^{1/3}$ and $A^\mathrm{PN}_{\ell m}(f)$ is the Pad\'e resummed version of the Fourier domain 3PN amplitude of the 32 and 43 modes. The Fourier domain amplitude is obtained using the stationary phase approximation as in~\cite{VanDenBroeck:2006qu}, starting from the time-domain PN results in~\cite{Blanchet:2008je}. We use $P^{0}_{4}$ and $P^{0}_{3}$ Pad\'e approximants for the $32$ and $43$ modes, respectively, similar to our treatment of the other modes in~\cite{Mehta:2017jpq}.

The modeling of $A_{\ell m}^\mathrm{RD}(f)$ exactly follows Eq.~(2.10) of \cite{Mehta:2017jpq}, i.e.,
\begin{equation}
A_{\ell m}^\mathrm{RD}(f) =  w_{\ell m} ~ e^{-\lambda_{\ell m}} ~ \left|\B_{\ell m}(f) \right|,
\label{eq:phenom_ampl_model_RD}
\end{equation}
where $\B_{\ell m}(f)$ is the Fourier transform of a damped sinusoid:  
\begin{equation}
\B_{\ell m}(f) = \frac{\sigmaqnm_{\ell m} - \mathrm{i} \, f}{\fqnm_{\ell m}^2 + (\sigmaqnm_{\ell m} - \mathrm{i} \, f)^2}. 
\end{equation}
The frequencies  $\fqnm_{\ell m}$  and $\sigmaqnm_{\ell m}$ are the real and imaginary parts of the $\ell m 0$ quasi-normal mode frequency of a Kerr black hole $\Omega_{\ell m 0} = 2\pi \, (\fqnm_{\ell m} + \mathrm{i} \, \sigmaqnm_{\ell m})$, determined from the mass and spin of the final black hole using the fits from~\cite{Berti_ringdown,Berti:2005ys,Berti:2009kk}. The phenomenological parameter $\lambda_{\ell m}$ in Eq.~\eqref{eq:phenom_ampl_model_RD} is determined from fits to numerical Fourier transforms of the hybrid waveforms, while $w_{\ell m}$ is a normalization constant to make the amplitudes continuous at the merger-ringdown matching frequency $f^\mathrm{A}_{\ell m}$. The mass and spin of the final black hole are computed from the masses of the initial black holes using fitting formulae calibrated to NR simulations, given in~\cite{Pan:2011gk}. 

Similarly, for the phase model we have 
\begin{equation}
\Psi^\text{U, mod}_{\ell m}(f) = \begin{dcases}
\Psi_{\ell m}^{\mathrm{IM}}(f),& \,\, f< f^{\rm P}_{\ell m},\\
\Psi_{\ell m}^{\mathrm{RD}}(f),& \,\, f \geq f^{\rm P}_{\ell m},
\end{dcases}
\label{eq:phenom_phase_model}
\end{equation}
\begin{equation}
\Psi^{\rm IM}_{\ell m}(f) = \Psi^{\rm PN}_{\ell m} (f) + \sum_{k=0}^{k=3}(\beta_{k, \, \ell m}+\beta_{k, \, \ell m}^\mathrm{L}\,\ln v_f + \beta_{k, \, \ell m}^\mathrm{L2}\,\ln^2 v_f)\,v_f^{k+8},
\label{eq:IMphase_model}
\end{equation}
where $\Psi^{\rm PN}_{\ell m} (f)$ is the PN phasing of the $\ell m$ mode and $f^{\rm P}_{\ell m}$ denotes the transition frequency from the inspiral-merger part of the waveform to the ringdown in the phase. The ringdown  part of the phase is modelled as
\be
\label{eq:ringdown_phase_model}
\Psi^{\rm RD}_{\ell m}(f) = 2 \pi f t^\mathrm{P}_{\ell m} + \phi^\mathrm{P}_{\ell m} +  \arctan \, \B_{\ell m}(f),
\ee
where $t^\mathrm{P}_{\ell m}$ and $\phi^\mathrm{P}_{\ell m}$ are computed by matching two phases ($\Psi^{\rm IM}_{\ell m}$ and $\Psi^{\rm RD}_{\ell m}$) and their first derivative at the matching frequency $f^\mathrm{P}_{\ell m}$.

Now, the phenomenological parameters appearing in the analytical models (for the $32$ and $43$ modes) are represented as quadratic functions of the symmetric mass ratio $\eta$:
\begin{equation}
\begin{split}
\label{eq:phenfits_phase_fits}
\alpha_{i, \, \ell m} &= a_{i, \, \ell m}^{\alpha} + b_{i, \, \ell m}^{\alpha}\,\eta + c_{i, \, \ell m}^{\alpha}\,\eta^2\,,\\
\alpha_{i, \, \ell m}^{\rm L} &= a_{i, \, \ell m}^{\alpha, \rm L} + b_{i, \, \ell m}^{\alpha,\rm L} \,\eta + c_{i, \, \ell m}^{\alpha, \rm L}\,\eta^2\,,\\
\beta_{k, \, \ell m} &= a_{k, \, \ell m}^{\beta} + b_{k, \, \ell m}^{\beta}\,\eta + c_{k, \, \ell m}^{\beta}\,\eta^2\,,\\
\beta_{k, \, \ell m}^{\rm L} &= a_{k, \, \ell m}^{\beta,\rm L} + b_{k, \, \ell m}^{\beta,\rm L} \,\eta + c_{k, \, \ell m}^{\beta,\rm L}\,\eta^2\,,\\
\beta_{0, \, \ell m}^{\rm L2} &= a_{0, \, \ell m}^{\beta,\rm L2} + b_{0, \, \ell m}^{\beta,\rm L2} \,\eta + c_{0, \, \ell m}^{\beta,\rm L2}\,\eta^2\,,\\
\lambda_{\ell m}  &=   a_{\ell m}^{\rm \lambda} + b_{\ell m}^{\rm  \lambda} \,\eta + c_{\ell m}^{\rm  \lambda}\,\eta^2\,, \\ 
f^{X}_{\ell m}  &=  ( a_{\ell m}^{X} + b_{\ell m}^{X} \,\eta + c_{\ell m}^{X}\,\eta^2)\, / M,
\end{split}
\end{equation}
where the index $i$ runs from 0 to 1 and $k$ runs from 0 to 3, while $X\in\{A,P\}$. 
We also refit the phase of the $22$ mode using the smaller number of coefficients given in~\eqref{eq:IMphase_model}; the fit in~\cite{Mehta:2017jpq} has the same form, except that the sum extends up to $k = 4$ instead of $k = 3$. We use this refit since it improves the $22$ mode's overlap with high mass ratio hybrid waveforms. Figure~\ref{fig:comp_waveform} provides a comparison of the amplitudes  and phases of the unmixed modes in the Fourier domain with the analytical fits given
by Eqs.~(\ref{eq:phenom_ampl_model}) and~(\ref{eq:phenom_phase_model}). Figure~\ref{fig:coef_fit} shows the values of the phenomenological parameters estimated from the hybrid waveforms, as well as the fits given in Eq.~(\ref{eq:phenfits_phase_fits}).\footnote{We find that quadratic polynomials in $\eta$ provide sufficiently accurate fits in terms of mismatches. Hence we do not consider higher order fits, even though there appears to be some substructure that would require a higher-order polynomial to fit (Fig.~\ref{fig:coef_fit}). It is possible that some of the structure seen in the $32$ and $43$ modes' coefficients for $\eta$ close to $0.25$ is related to the fact that the mode mixing removal does not work as well for $q < 3$, as discussed below.} 

\begin{figure}[htb] 
	\includegraphics[width=3.4in]{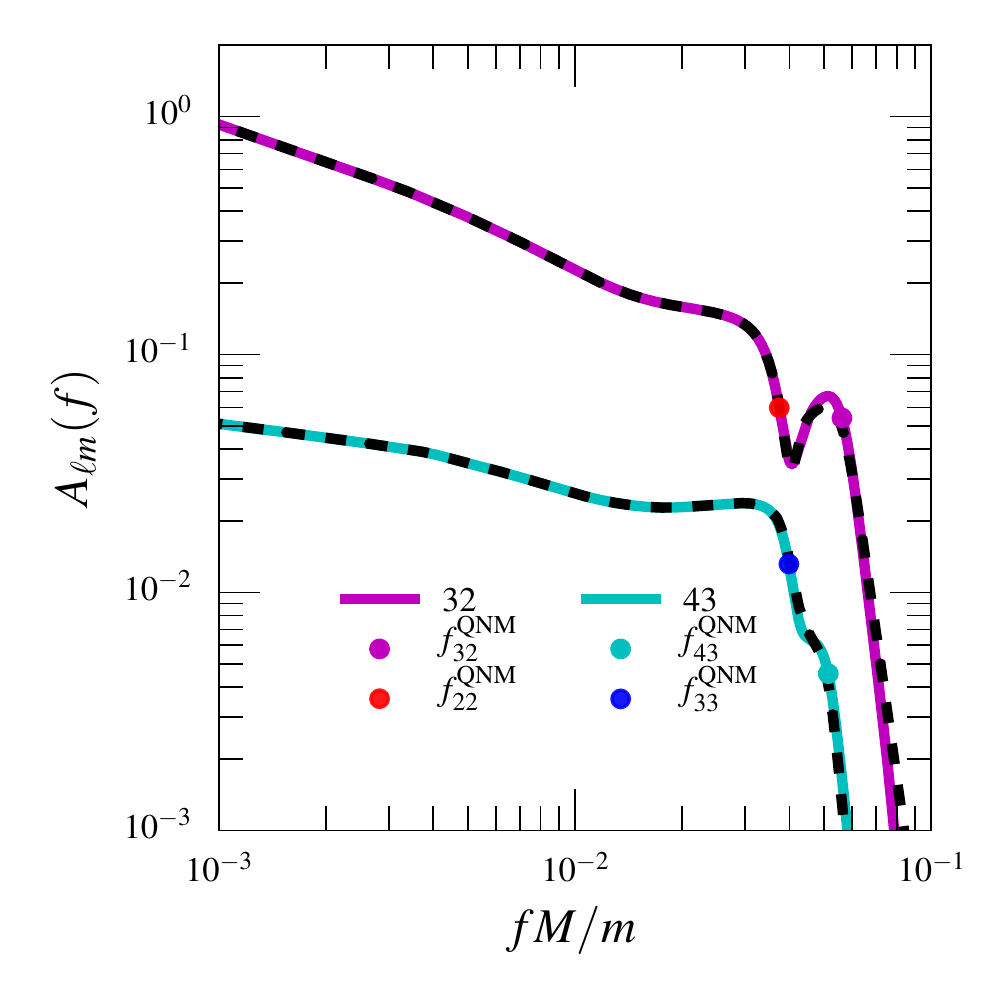}
	\vspace*{-2mm}
	\caption{Comparison of the amplitude of the mixed modes for a mass ratio $q = 4$, showing the hybrid (solid lines) and analytical model (dashed lines).}
	\label{fig:mixed_modes_comp}
\end{figure}

\subsection{Adding the mode mixing contribution into unmixed  modes}

Having constructed analytical models for the amplitude and phase of unmixed modes, we need to add the mode mixing contribution back into this model in order to get the analytical model for the amplitude and phase of the spherical harmonic modes ${}^\text{Y}\tilde{\h}_{\ell m}^{\mathrm{R}}(f)$. This is done as follows: We denote the (Fourier domain) model waveform by
\begin{equation}
{}^\text{Y}\tilde{\h}^\text{mod}_{\ell m}(f) := A^\text{U, mod}_{\ell m}(f) ~ e^{\mathrm{i} \, \Psi^\text{U, mod}_{\ell m}(f)},
\end{equation}
where $A^\text{U, mod}_{\ell m}(f)$ and $\Psi^\text{U, mod}_{\ell m}(f)$ are given by Eqs.~\eqref{eq:phenom_ampl_model} and~\eqref{eq:phenom_phase_model}. We then write
\begin{equation}
\begin{split}
{}^\text{M}\tilde{\h}_{32}^{\mathrm{R}, \text{ mod}}(f) &= {}^\text{Y}\tilde{\h}_{22}^{\mathrm{R}, \text{ mod}}(f)\rho_{2320}^{\mathrm{R}} - {}^\text{Y}\tilde{\h}_{22}^{\mathrm{I}, \text{ mod}}(f)\rho_{2320}^{\mathrm{I}}\\
&\quad + {}^\text{U}\tilde{\h}_{320}^{\mathrm{R}, \text{ mod}}(f)\mu_{2330}^{\mathrm{R}} - {}^\text{U}\tilde{\h}_{320}^{\mathrm{I}, \text{ mod}}(f)\mu_{2330}^{\mathrm{I}},
\end{split}
\label{eq:add_mode_mixing}
\end{equation}
where 
\begin{equation}
{}^\text{U}\tilde{\h}^\text{mod}_{320}(f) := \varepsilon_{32} {}^\text{Y}\tilde{\h}^\text{mod}_{\ell m}(f)  ~ e^{\mathrm{i} \, \delta_{32}\pi }.
\end{equation}
The expressions for the $43$ mode are analogous.
%
%
Here  we have introduced two free parameters, $\varepsilon_{\ell m}$ and $\delta_{\ell m}$, corresponding to the amplitude ratio and phase difference at $f_{\ell m}^\text{mix}$. We fit these parameters by minimizing the mismatch of ${}^\text{M}\tilde{\h}_{32}^{\mathrm{R}}(f)$ with the corresponding hybrid mode.
They are represented as cubic functions of the symmetric mass ratio; 
we find similar functional behavior when we compute the amplitude ratio and phase difference between mixed (spherical) and spheroidal modes [Eq.~\eqref{eq:freq_mod_remov}] at $f_{\ell m}^\text{mix}$. Specifically,
 \begin{eqnarray}
 \varepsilon_{\ell m} &=& a^{\varepsilon}_{ \ell m} + b^{\varepsilon}_{ \ell m}\,\eta + c^{\varepsilon}_{ \ell m}\,\eta^2\ + d^{\varepsilon}_{\ell m}\,\eta^3, \nonumber \\
 \delta_{\ell m} &=& a^{\delta}_{ \ell m} +  b^{\delta}_{ \ell m}\,\eta +  c^{\delta}_{ \ell m}\,\eta^2\ + d^{\delta}_{\ell m}\,\eta^3, \nonumber \\
 \end{eqnarray}
 where $\ell m \in \{32,43\}$. The fits for parameters $\varepsilon_{\ell m}$ and $\delta_{\ell m}$ are shown in the Fig.~\ref{fig:coef_fit}.

\begin{figure*}[htb] \begin{center}
		\includegraphics[width=5.5in]{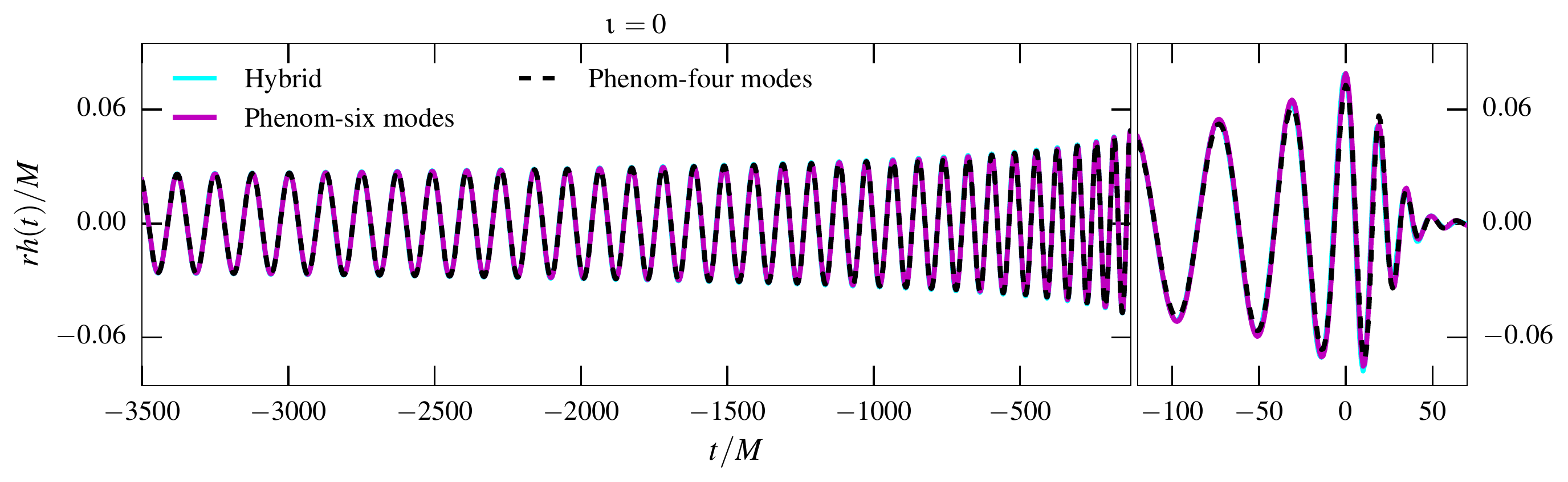}\\
		\includegraphics[width=5.5in]{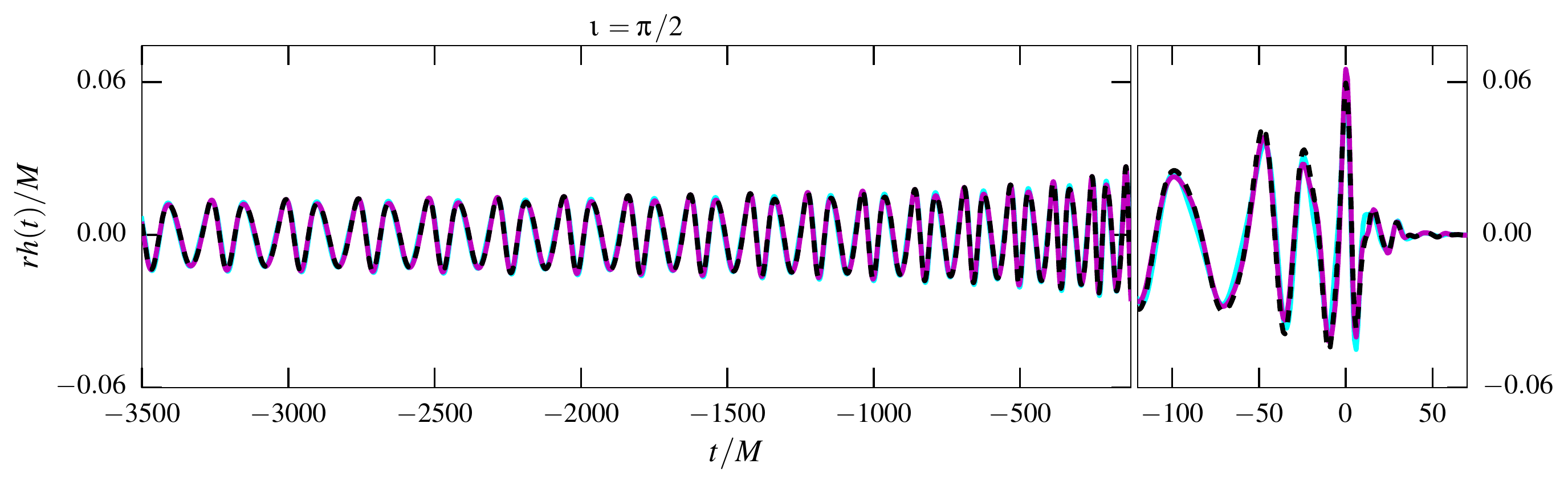}
		\caption{Comparison between hybrid waveforms and our analytical phenomenological waveforms for a binary with mass ratio $q = 10$. Hybrid waveforms are constructed using all the modes with $\ell \leq 4$, except the $m = 0$ modes. Phenomenological waveforms are constructed by taking the (discrete) inverse Fourier transform of the analytical model waveforms in the Fourier domain. The top panel corresponds to a ``face-on'' binary (inclination angle $\iota=0$) while the bottom panel corresponds to an ``edge-on'' binary ($\iota=\pi/2$). The two phenomenological waveforms correspond to the current model with and without the $32$ and $43$ modes.
		}
		\label{fig:hybrid_phenom_comparison_td}
\end{center} \end{figure*}
 
 \begin{figure*}[htb] \begin{center}
 		\includegraphics[width=7in]{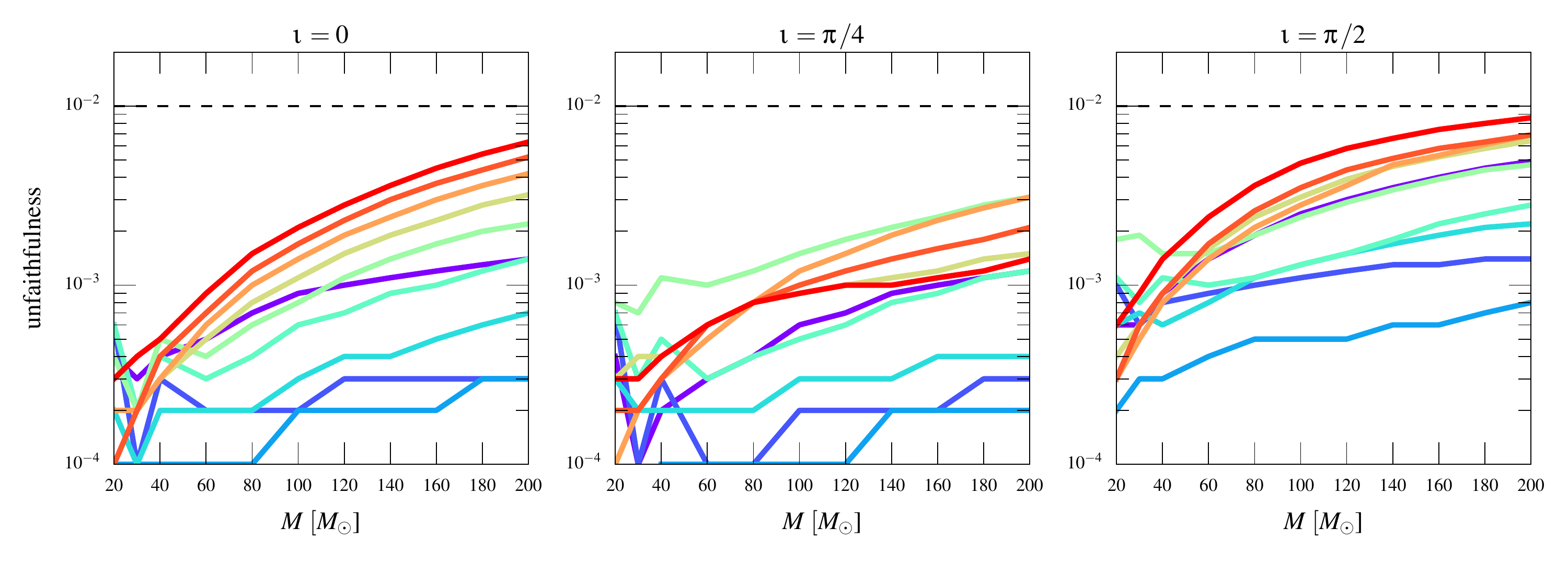}
 		\includegraphics[width=7in]{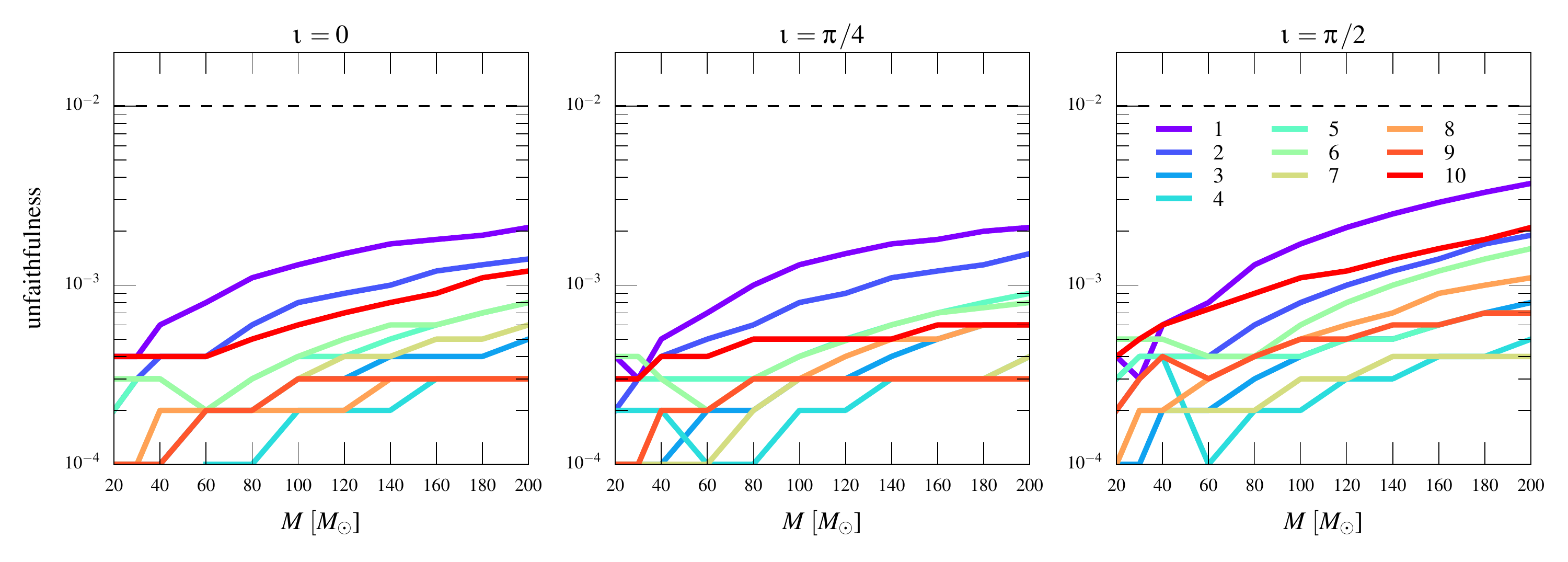}
 		\vspace*{-2mm}
 		\caption{The unfaithfulness (mismatch)  of the analytical model waveform family towards hybrid waveforms for inclination angle $\iota =\pi/2$. The analytical model waveform family in the top panel contains only the $22$, $33$, $44$, and $21$ modes while in the bottom panel the mixed modes we model here are also included, i.e., also the $32$ and $43$ modes. The horizontal axes report the total mass of the binary and different curves correspond to different mass ratios $q$ (shown in the legend). Horizontal black dashed lines correspond to a mismatch of 1\%. The overlaps are computed assuming the design power spectrum of Advanced LIGO (in the ``high-power, zero-detuning'' configuration~\cite{adligo-psd}), assuming a low-frequency cutoff of 20 Hz. We do not consider a smaller low-frequency cutoff or smaller total masses due to computational difficulties with constructing hybrid waveforms starting from lower dimensionless frequencies.
 		\label{fig:mismatch_allmodes}
		}
 \end{center} \end{figure*}
 
The amplitude and phase of the complete model for ${}^\text{Y}\tilde{\h}_{\ell m}^{\mathrm{R}}(f)$ for the mixed modes ($32$ and $43$) are finally constructed as follows:
\begin{subequations}
\begin{align}
A_{\ell m}(f) &= \begin{dcases}
|{}^\text{Y}\tilde{\h}_{\ell m}^{\mathrm{R}, \text{ mod}}(f)|,& \,\, f< f_{\ell m}^\text{mix},\\ 
w_{\ell m}^\text{M}|{}^\text{M}\tilde{\h}_{\ell m}^{\mathrm{R}, \text{ mod}}(f)|,& \,\, f \geq f_{\ell m}^\text{mix},
\end{dcases}\\
\Psi_{\ell m}(f) &= \begin{dcases}
\arg({}^\text{Y}\tilde{\h}_{\ell m}^{\mathrm{R}, \text{ mod}}(f)),& \,\, f< f_{\ell m}^\text{mix},\\  
\phi_{\ell m}^\text{M} + \arg({}^\text{M}\tilde{\h}_{\ell m}^{\mathrm{R}, \text{ mod}}(f)),& \,\, f \geq f_{\ell m}^\text{mix}.
\end{dcases}
\end{align}
\end{subequations}
The parameters $w_{\ell m}^\text{M}$ and $\phi_{\ell m}^\text{M}$ ensure the continuity of amplitude and phase at $f_{\ell m}^\text{mix}$, respectively. We compare the results of the final model for the spherical harmonics with the hybrids in Fig.~\ref{fig:mixed_modes_comp}.

So far, we have used $q = 4$ for all our illustrations. We chose this mass ratio to give a clean illustration in a case where the higher modes are relatively prominent and the mode mixing is still fairly large. (The mode mixing decreases as the mass ratio increases for nonspinning binary black holes, as the final spin decreases with increasing mass ratio.) We find that the mode mixing removal is less effective for smaller mass ratios, for reasons that we do not understand. Nevertheless, we still find that our model provides an accurate representation of the waveforms in these cases, as is shown by the match calculations below. We give illustrations of the mode mixing removal and the accuracy of the model for $q = 2.32$ in Appendix~\ref{sec:lower_mass_rat_plot}.

\subsection{Assessing the accuracy of the analytical model}
\label{sec:waveform_accuracy}
We assess the accuracy of our model by computing mismatches with the same set of $10$ hybrid waveforms used to validate the model in~\cite{Mehta:2017jpq} (which only share $4$ waveforms---primarily high mass-ratio ones---with the set of $8$ waveforms that are used to construct the model; see Appendix~\ref{sec:SXS}). The overlaps are computed assuming the design power spectrum of Advanced LIGO (in the ``high-power, zero-detuning'' configuration~\cite{adligo-psd}),\footnote{This noise curve has recently been updated slightly with newer predictions for the thermal noise~\cite{updated_adligo-psd}. We use the older version.} assuming a low-frequency cutoff of 20 Hz, for a range of total masses.

Figure~\ref{fig:hybrid_phenom_comparison_td} shows the comparison of our waveform model in the time domain against the hybrid waveforms for two cases, firstly when the model waveforms contain only four modes, i.e., $22$, $33$, $44$, and $21$, and secondly when it also includes the $32$ and $43$ modes in addition to the four modes mentioned before. The hybrid waveforms contain all modes with $\ell \leq 4$, except for the $m = 0$ modes, which are small and not well-resolved in the NR simulations. We see that the inclusion of the two additional modes improves the agreement between the hybrid and phenomenological waveforms. Additionally, comparing with Fig.~3 in~\cite{Mehta:2017jpq}, we see the improvement in the face-on case due to the refit of the $22$ mode, as well as the inclusion of the $32$ mode; the $43$ mode does not contribute for a face-on binary.

While Fig.~\ref{fig:hybrid_phenom_comparison_td} only shows qualitative agreement between the phenomenological and hybrid waveforms, Fig.~\ref{fig:mismatch_allmodes} shows the mismatch plots.  The top panel plots show the mismatch (unfaithfulness) between the hybrid waveforms and the case where the model waveforms contain only four modes for various inclination angles. The bottom panel plots show the mismatch after including the  $32$ and $43$ modes in the model waveforms for the same inclination angles. We see that for high mass ratio waveforms, the maximum mismatch reduces  from $1\%$ to $0.2\%$ for the highest inclination angle $\iota=\pi/2$. However, mismatches are even lower ($\sim 0.05\%$) for other inclination angles. The lower mass ratio cases are almost unaffected, though they show  a little improvement. This is expected, because the contribution of higher modes is significant for high mass ratio and inclination angles.

\begin{figure*}[htb] \begin{center}
		\includegraphics[height=3.0in]{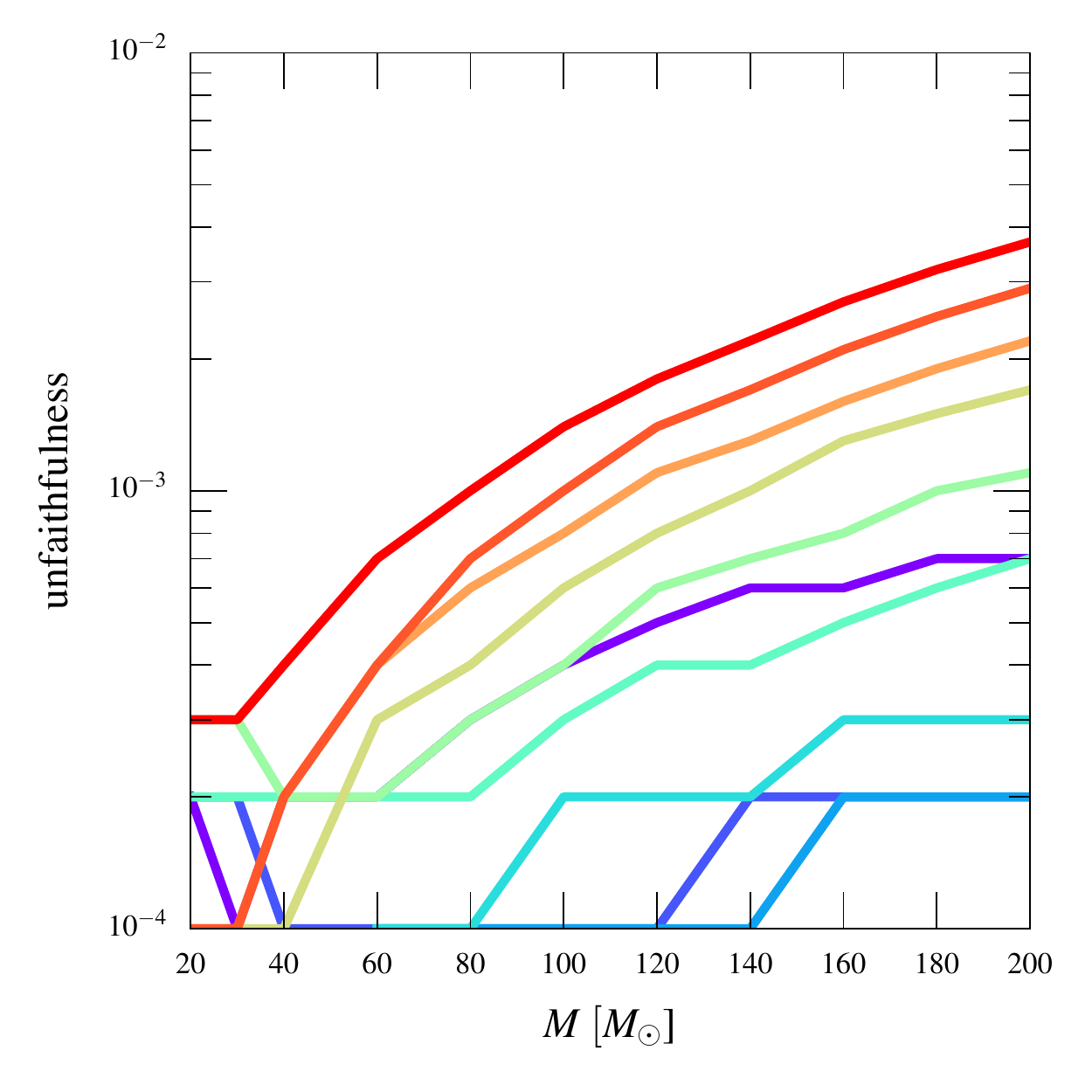}
		\includegraphics[height=3.0in]{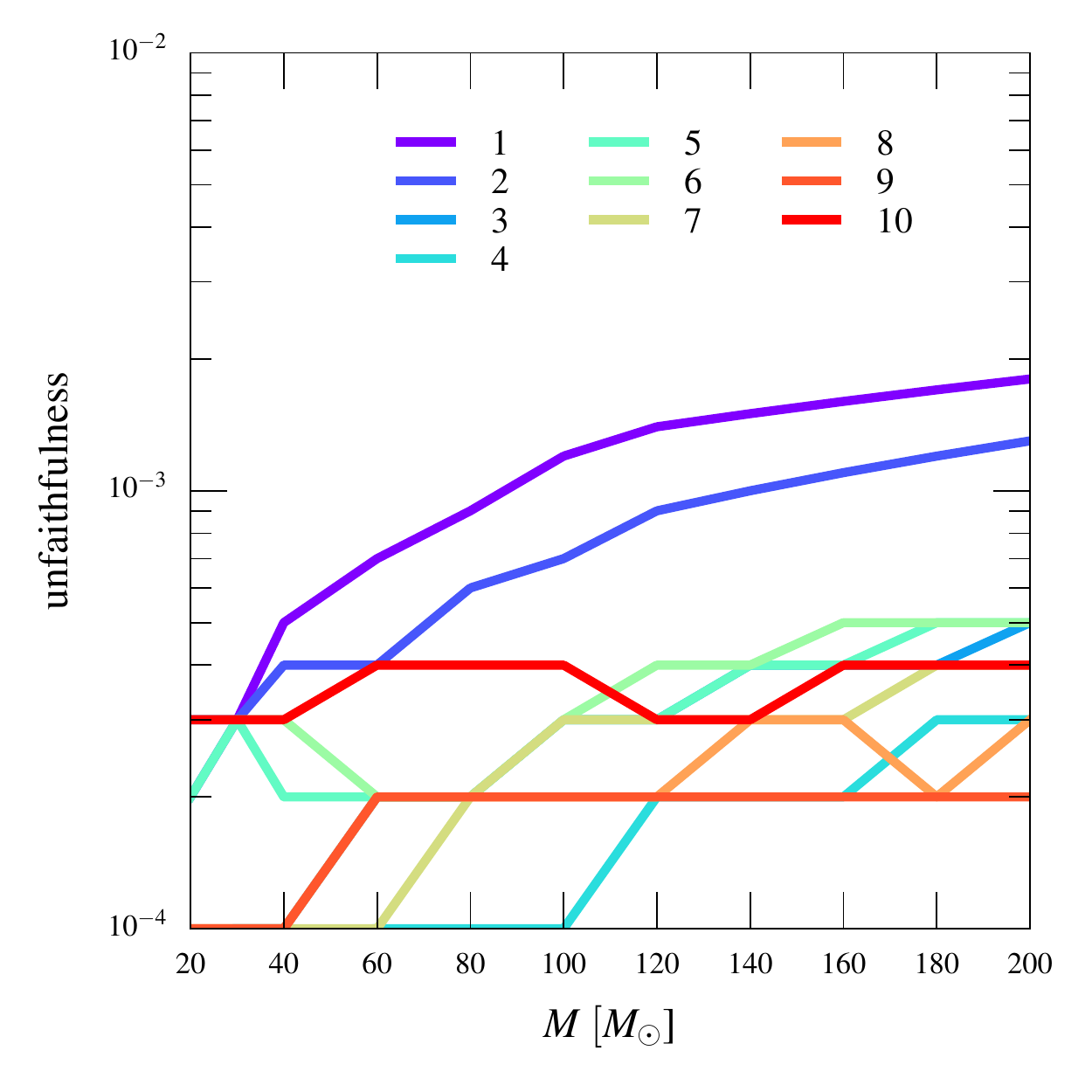}
		\vspace*{-2mm}
		\caption{The unfaithfulness (mismatch) of the analytical model waveform $22$ mode  against the hybrid $22$ mode. The left plot shows the mismatch for the previous phenomenological $22$ mode and right plot shows the mismatch for the current phenomenological $22$ mode which has been remodeled. The horizontal axes report the total mass of the binary and different curves correspond to different mass ratios $q$ (shown in the legend). We can see a significant improvement in the mismatch for high mass ratio waveforms in the right-hand plot. 
		}
		\label{fig:mismatch_mixed_modes}
\end{center} \end{figure*}

We also show the improvement in the accuracy of the model for the 22 mode alone for higher mass ratios in Fig.~\ref{fig:mismatch_mixed_modes}. This comes from a refit of this mode's phase. Unfortunately, this improvement for higher mass ratios comes at the cost of a somewhat larger mismatch for mass ratios of $1$ and $2$. Future work will consider improvements to the structure of the model to improve the mismatch for small mass ratios.

\section{Summary and Conclusions}
\label{sec:summ_concl}
In this paper, we extend our analytical frequency-domain phenomenological higher-mode model for gravitational waveforms from nonspinning binary black holes~\cite{Mehta:2017jpq} to include two additional subdominant modes, namely the $32$ and $43$ modes, in addition to the $22$, $33$, $44$, and $21$ modes. This waveform family now has a faithfulness of $> 99.6\%$ for binaries up to a mass ratio of $10$ and a total mass of $200M_\odot$, using the design sensitivity Advanced LIGO noise curve and a low-frequency cutoff of $20$~Hz. The two additional modes that we model in this paper, i.e., $32$ and $43$, exhibit the effects of mode-mixing, i.e., having multiple spheroidal harmonic ringdown modes mixed into a single spherical harmonic mode. This leads to bumps in the ringdown part of the waveform. We have introduced a simple way of approximately extracting the unmixed (spheroidal harmonic) modes. We then model these unmixed modes using the method used for the other modes in~\cite{Mehta:2017jpq}. We then reinstate the mode mixing using the models for the unmixed modes to obtain the final model for the spherical harmonic modes. We also refit our model for the dominant $22$ mode to improve its accuracy for large mass ratios. This is the first analytical inspiral-merger-ringdown waveform family that models the mode mixing effect. 

We note that efforts are underway to construct effective-one-body/phenomenological models for binary black holes with non-precessing spins~\cite{London:2017bcn, Cotesta:2018fcv}. The simple prescription that we use to model mode mixing effects may be used in these models as well. Future models aiming for a highly accurate description of non-quadrupole modes may also need to consider other sources of mode mixing, e.g., the mode-mixing due to boosts and displacements from the origin, discussed in~\cite{Boyle:2015nqa}. Additionally, as discussed in Appendix~\ref{sec:lower_mass_rat_plot}, more accurate determination of the waveforms at infinity in numerical simulations, e.g., through Cauchy-characteristic extraction~\cite{Reisswig:2009us,Taylor:2013zia,Chu:2015kft,Handmer:2016mls}, will likely be necessary input for precise models. As GW observations are becoming precision probes of physics and astrophysics, accuracy requirements on GW templates can only grow. 


\acknowledgments

We are grateful to the SXS collaboration for making a public catalog of numerical-relativity waveforms. We also thank Emanuele Berti and Michael Boyle for useful discussions and clarifications, Frank Ohme for helpful comments, and Mark Scheel for providing a Cauchy-characteristic extraction SXS waveform for comparison. A.~K.~M., P.~A., and V.~V.\ acknowledge support from the Indo-US Centre for the Exploration of Extreme Gravity funded by the Indo-US Science and Technology Forum (IUSSTF/JC-029/2016). N.~K.~J.-M.\ acknowledges support from the AIRBUS Group Corporate Foundation through a chair in ``Mathematics of Complex Systems'' at the International Centre for Theoretical Sciences (ICTS) and from STFC Consolidator Grant No.~ST/L000636/1. Also, this work has received funding from the European Union's Horizon 2020 research and innovation programme under the Marie Sk{\l}odowska-Curie Grant Agreement No.~690904. P.~A.'s research was supported by the Max Planck Society through a Max Planck Partner Group at ICTS-TIFR, and by the Canadian Institute for Advanced Research through the CIFAR Azrieli Global Scholars program. V.~V.'s research was supported by the Sherman Fairchild Foundation, and NSF grants PHY--170212 and PHY--1708213 at Caltech. Computations were performed at the ICTS cluster Alice. This document has LIGO preprint number LIGO-P1800203-v6.

\appendix

\section{Numerical relativity waveforms}
\label{sec:SXS}
In Table~\ref{tab:NR_waveforms}, we list the SXS waveforms that are used to fit the coefficients appearing in Sec.~\ref{sec:phenom_model} and to assess the accuracy of our model as described in Sec.~\ref{sec:waveform_accuracy}.

\begin{table}
	\centering
	\begin{tabular}{c@{\quad} c@{\quad}c@{\quad}c@{\quad}r}
		\toprule
		Simulation ID & $q$ & $M\omega_\mathrm{orb}$ & $e$ & \# orbits \\
		\midrule
		\emph{Fitting} \\ 
		\midrule
		SXS:BBH:0198 & $1.20$  &  $0.015$ &  $2.0 \times 10^{-4}$ &  $20.7$\\
		SXS:BBH:0201 & $2.32$ &  $0.016$ &  $1.4 \times 10^{-4}$ &  $20.0$ \\
		SXS:BBH:0200 & $3.27$ &  $0.017$ &  $4.1 \times 10^{-4}$ &  $20.1$ \\
		SXS:BBH:0182 & $4.00$ &  $0.020$ &  $6.8 \times 10^{-5}$ &  $15.6$\\
		SXS:BBH:0297 & $6.50$ & $0.021$ &  $5.9 \times 10^{-5}$ &  $19.7$\\
		SXS:BBH:0063 & $8.00$ &  $0.019$ &  $2.8 \times 10^{-4}$ &  $25.8$ \\
		SXS:BBH:0301 & $9.00$ &  $0.023$ &  $5.7 \times 10^{-5}$ &  $18.9$ \\
		SXS:BBH:0185 & $9.99$ &  $0.021$ &  $2.9 \times 10^{-4}$ &  $24.9$ \\
		\midrule
		\emph{Verification}\\ 
		\midrule
		SXS:BBH:0066 & $1.00$ &  $0.012$ &  $6.4 \times 10^{-5}$ &  $28.1$\\
		SXS:BBH:0184 & $2.00$ &  $0.018$ &  $7.6 \times 10^{-5}$ &  $15.6$\\
		SXS:BBH:0183 & $3.00$ &  $0.019$ &  $6.3 \times 10^{-5}$ &  $15.6$\\
		SXS:BBH:0182 & $4.00$ &  $0.020$ &  $6.8 \times 10^{-5}$ &  $15.6$\\
		SXS:BBH:0187 & $5.04$ &  $0.019$ &  $5.0 \times 10^{-5}$ &  $19.2$\\
		SXS:BBH:0181 & $6.00$ &  $0.017$ &  $7.9 \times 10^{-5}$ &  $26.5$\\
		SXS:BBH:0298 & $7.00$ &  $0.021$ &  $4.0 \times 10^{-4}$ &  $19.7$\\
		SXS:BBH:0063 & $8.00$ &  $0.019$ &  $2.8 \times 10^{-4}$ &  $25.8$ \\
		SXS:BBH:0301 & $9.00$ &  $0.023$ &  $5.7 \times 10^{-5}$ &  $18.9$ \\
		SXS:BBH:0185 & $9.99$ &  $0.021$ &  $2.9 \times 10^{-4}$ &  $24.9$ \\
		\bottomrule
	\end{tabular}
	\caption{Summary of the parameters of the NR waveforms used in this paper: $q
		:= m_1/m_2$ is the mass ratio of the binary, $M \omega_\mathrm{orb}$ is the
		orbital frequency after the junk radiation and $e$ is the residual
		eccentricity. The waveforms listed under the title \emph{Fitting} are used to 
		produce the analytical fits described in Sec.~\ref{sec:phenom_model} while
		those listed under the title \emph{Verification} are used for assessing the
		faithfulness of the analytical model in Sec.~\ref{sec:waveform_accuracy}.}
	\label{tab:NR_waveforms}
\end{table}

\section{Mode mixing removal for lower mass ratios}
\label{sec:lower_mass_rat_plot}

\begin{figure*}[htb] \begin{center}
		\includegraphics[height=3.0in]{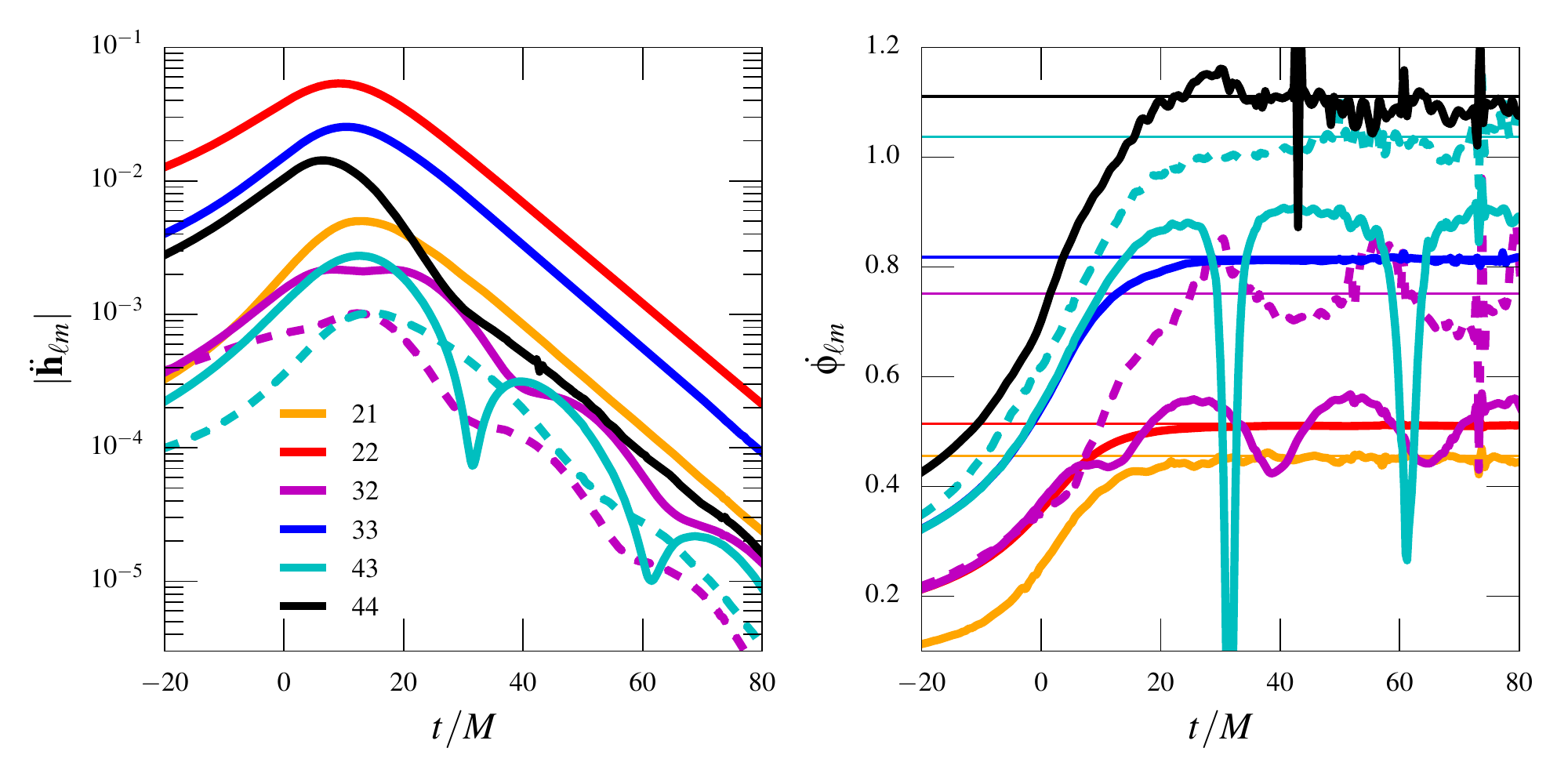}\\
		\vspace*{-2mm}
		\caption{This is the analog of Fig.~\ref{fig:amp_comp} for mass ratio $q=2.32$. The left-hand plot shows the mode mixing removal in the time domain amplitude of the second time derivatives of the modes and the right-hand plot shows the effects of the mode mixing removal on the instantaneous frequency of the second time derivatives of the modes. The solid lines show the spherical harmonic modes and the dashed lines show the unmixed spheroidal harmonic $320$ and $430$ modes constructed using the procedure in Sec.~\ref{sec:mode_mixing_removal}. We see that there are considerably larger oscillations in the instantaneous frequency of the $320$ mode than in Fig.~\ref{fig:amp_comp}, particularly in the frequency. We also see some numerical noise in the frequency plots, which we find can be attributed to the extrapolation procedure used to obtain the waveform at infinity.} 
		\label{fig:mixed_modes_low_mass_ratio}
\end{center} \end{figure*}
\begin{figure}[h]
		\includegraphics[height=3.0in]{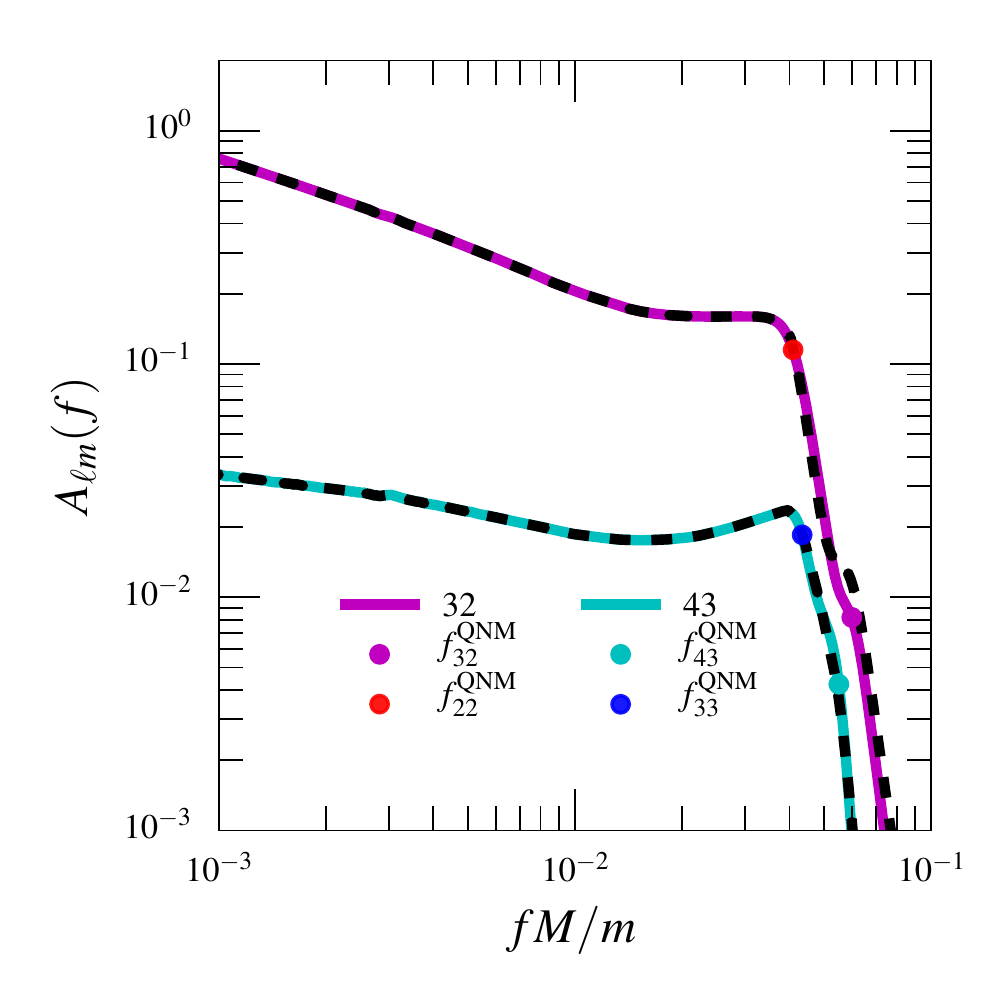}
		\vspace*{-2mm}
		\caption{This is the analog of Fig.~\ref{fig:mixed_modes_comp} for mass ratio $q=2.32$. The solid lines show the amplitude of the hybrid and the dashed lines show the amplitude of our analytical model for the mixed modes (32 and 43). We can see that the model is able to reproduce the hybrid modes quite well, even though our mode-mixing removal method for this mass ratio is less effective as compared to the same for higher mass ratios.}
		\label{fig:mixed_modes_comp_low_mass_rat}
\end{figure}

We give analogs of Figs.~\ref{fig:amp_comp} and~\ref{fig:mixed_modes_comp} for a mass ratio of $q = 2.32$ in Figs.~\ref{fig:mixed_modes_low_mass_ratio} and~\ref{fig:mixed_modes_comp_low_mass_rat}. The first of these figures illustrates that the mode mixing removal is still effective in improving the agreement of the instantaneous frequency with the expected QNM frequency, and in reducing the amplitude oscillations of the 32 mode. However, the mode mixing removal is less effective for $q \lesssim 3$, for reasons we do not fully understand. Nevertheless, the second figure shows that the final model for the mixed modes still agrees well with the Fourier transform of the hybrid.

The numerical noise we find in the instantaneous frequency of the modes is reduced when considering the NR waveform with no extrapolation to infinity. Experimentation with the equal-mass nonspinning Cauchy-characteristic extraction SXS waveform from~\cite{Taylor:2013zia} finds that this does not suffer from the numerical noise that is present in the instantaneous frequencies of the analogous finite radius or extrapolated equal-mass nonspinning SXS waveform modes.

\newpage
\bibliography{ModeMixing}

\end{document}